\documentclass[12pt]{extarticle}
\usepackage[utf8]{inputenc}
\usepackage[english]{babel}
\usepackage{fontenc}
\usepackage{amsmath}
\usepackage{mathtools}
\usepackage{amssymb}
\usepackage{amsthm}
\usepackage{xcolor}
\usepackage{enumitem}
\usepackage{appendix}
\usepackage{graphicx}
\usepackage{booktabs}
\usepackage{float}
\usepackage{multirow}
\usepackage{geometry}

\newtheoremstyle{nonitalic}%
  {}%
  {}%
  {\normalfont}%
  {}%
  {\bfseries}%
  {.}%
  { }%
  {}%

\theoremstyle{nonitalic}

\newtheorem{remark}{Remark}

\theoremstyle{plain}
\newtheorem{lemma}{Lemma}
\newtheorem{thrm}{Theorem}

\newtheorem{corollary}{Corollary}

\title{Necessary Conditions and Existence of Optimal Execution in a Limit Order Book Dynamics Model Governed by a First-Order Partial Differential Equation}
\author{A.O. Belyakov, \\
Moscow School of Economics, Lomonosov Moscow State University\\
National University of Science and Technology \lq{}\lq{} MISiS\rq{}\rq{}, Moscow\\
\\ 
A.S. Platov,\\
National University of Science and Technology \lq{}\lq{}MISiS\rq{}\rq{}, \\
Vega Institute Foundation
}

\begin{document}

\maketitle

\begin{abstract}
We study the problem of optimal execution of a fixed purchase volume in a limit order book market with bounded execution rate. Initially arbitrary order density is governed by a first-order hyperbolic transport equation, where the velocity of order movement depends on the difference between the current market price and the long-term fundamental price with a drift. 
We reduce the infinite-dimensional optimal control problem with a homogeneous partial differential equation constraint to a finite-dimensional one using the method of characteristics. That allows using necessary optimality conditions in the form of the Pontryagin Maximum Principle to determine a switching function. We also prove the existence of an optimal control under natural assumptions on the initial order density. 
\end{abstract}

\section{Introduction}

This paper considers the problem of optimal execution of a fixed purchase volume in a limit order book (LOB) market. A limit order is understood as an instruction to sell an asset at a predetermined price. The aggregate of submitted limit orders forms the order book of an arbitrary initial shape as in \cite{AlfonsiFruthSchied2010}, but its density evolves over time.

Limit orders can continuously move in the price space, adapting to the current market situation. A key feature of the model is that this movement is described by a transport equation for the order density. It is assumed, following \cite{ObizhaevaWang2013}, that orders move towards the long-term asset price: if the current market price exceeds the long-term price (the market is overbought), orders shift downwards; otherwise, they shift upwards. We introduce a drift to the prices. The order density of arbitrary initial shape shifts and reacts to the buyer's actions, which makes the optimal execution problem highly nonlinear and distinguishes it from classical models with a uniform or self-similar order book.

Order execution occurs at the current market price level: all orders that reach this level are executed. The goal of this work is to investigate the problem of minimizing the total execution cost for a given purchase volume. The control is exerted through market price impact, while the control is subject to natural magnitude constraints. This excludes impulsive execution and necessitates a choice between regimes of maximum impact and waiting. Such a formulation leads to an optimal control problem for a system comprising a transport equation with a moving boundary and ordinary differential equations governing the market and the long-run prices.

In \cite{ObizhaevaWang2013}, a block order book model with exponential price recovery was proposed, where the optimal strategy combines initial and final block orders with continuous trading in between. In \cite{AlfonsiFruthSchied2010}, this model was extended to an arbitrary order book shape; explicit optimal strategies were obtained, and their robustness to the choice of the shape function was proven. However, in both works, the shape of the LOB and the recovery law are predefined.

The dynamics of the LOB, described by distributed systems, in particular parabolic reaction-diffusion equations, were considered in \cite{Donier2015}; a general overview of such models is given in \cite{Bouchaud2018}. In these models, the market price acts as a free boundary. Stochastic generalizations of such models were proposed in \cite{Hambly2020}. Optimal execution with risk considerations were studied in \cite{AlmgrenChriss2001} using static price impact functions, and a comprehensive framework for algorithmic trading with LOBs is detailed in \cite{CarteaJaimungalPenalva2015}.

An important distinction of our approach from the aforementioned models is that the order density dynamics is governed by a hyperbolic transport equation without sources or sinks, and the movement of orders is determined by the difference between the market and long-term prices with a drift. The equation is homogeneous, so the evolution of the book occurs not through the arrival of new orders or a predefined recovery process as in \cite{AlfonsiFruthSchied2010,ObizhaevaWang2013}, but through the directed displacement of already existing orders. Related approaches incorporating LOB shape into optimal execution can be found in \cite{GueantLehalleFernandezTapia2012}.

The main result of this work is the reduction of the infinite-dimensional optimal control problem to a finite-dimensional one using the method of characteristics, and the derivation of necessary optimality conditions in the form of the Pontryagin Maximum Principle.

\section{LOB Dynamics}

We describe the limit order book on the sell side (ask side) using a first-order partial differential equation:
\begin{equation}\label{eq:density}
    \partial_t n(p,t) + \partial_p (v(t)\, n(p,t)) = 0,\qquad n(p,0) = n_0(p)>0, \qquad p\geq A(t) > 0,\,t\in[0,T],
\end{equation}
where $n(p,t)$ is the unnormalized order density, i.e., the number of orders per unit price in the vicinity of the price level $p$ at time $t$; $n_0(p)$ is the initial distribution of orders across prices, with $n_0\in C^1([A(0),+\infty))$; and $A(t)$ is the current (market) price of the asset:
\begin{equation}\label{eq:market_price}
\frac{dA}{dt} = u -\rho\,(A - V)+b\,V,\qquad A(0)=A_0>0, \qquad \rho > 0,
\end{equation}
which, in the absence of buying pressure ($u\equiv 0$), converges at a rate $\rho$ to the long-term price $V$:
\begin{equation}\label{eq:f_price}
    \frac{dV}{dt} = \kappa\, u+b\,V,\quad V(0)=V_0>0, \quad 0<\kappa<1,
\end{equation}
where the coefficient $b$ accounts for the overall exponential growth (or decay) of prices.

We assume that the impact of the purchase, $u(t)\in[0,U]$, $u\in L^\infty([0,T])$, on the long-term price $V$ is strictly less ($\kappa<1$) than its impact on the current price $A$. The velocity of order movement across prices in \eqref{eq:density} is given by:
\begin{equation}\label{eq:velocity}
  v(t) = \gamma\, u(t)-\rho\,(A(t)-V(t))+b\,V(t),\qquad 0<\gamma<1.
\end{equation}
It differs from the rate of change of the market price $\frac{dA}{dt}$ in \eqref{eq:market_price} only by a lower sensitivity to the buying impact ($\gamma < 1$). Therefore, the characteristics of equation \eqref{eq:density} cross the market price trajectory only from top to bottom, and only when a purchase is executed ($u(t)>0$). In the absence of purchases ($u(t)=0$), the current price moves along the characteristics.

For simplicity, the right-hand side of the equation in \eqref{eq:density} is zero, which implies the absence of sources and sinks of sell orders. However, the support of their distribution density is unbounded from above in price, i.e., $n_0(p)>0$ for all $p\geq A_0 > 0$, meaning there are always sell orders at higher prices. We also assume for simplicity that the velocity $v(t)$ of order movement in the price space does not depend on the price $p$ of the current order, i.e., it is the same for all sellers.

Thanks to these simplifications, the solution to the equation can be obtained explicitly as follows.

\begin{lemma}\label{lm:lemma1}
    The density $n(p,t)$ is determined by the initial density $n_0$ at the price level 
    $$p - A(t) + A_0 + (1-\gamma)\int_0^t u(s)\, ds.$$
\end{lemma}
\begin{proof}
    The characteristic of \eqref{eq:density} with the initial value $p(0)=p_0\geq A_0$ is defined by the equation $\frac{dp}{dt} = v(t)$.
    Subtracting \eqref{eq:velocity} from \eqref{eq:market_price}, we obtain $\frac{dA}{dt}-\frac{dp}{dt} = (1-\gamma)u(t)$. Integrating this, we can write:
    $$
    A(t) - A_0 - p(t;p_0) + p_0 = \int_0^t (1-\gamma) u(s)\, ds.
    $$
    Since the order density along the characteristic remains constant:
    $$
    n(p(t;p_0), t) = n_0(p_0) = n_0\!\left( p(t) - A(t) + A_0 + (1-\gamma)\int_0^t u(s)\, ds \right).
    $$
\end{proof}

\begin{corollary}\label{co:nAt}
The density $n(p,t)$ at the current boundary $(A(t),t)$ is determined by the initial density at the price level $A_0+(1-\gamma)\int_0^t u(s)\, ds$.
\end{corollary}

\subsection{Purchase Flow}
Since the orders themselves possess an intrinsic displacement velocity $v(t)$, the flow of executed orders is determined by the relative velocity of the convergence between the price and the orders:
$$
\frac{dA}{dt}-v = (1-\gamma)\, u.
$$
Thus, the volume of executed orders over time $dt$ is described by the equation:
$$
dQ(t) = n(A(t),t)(1-\gamma) u(t)\, dt.
$$
\begin{remark}
    If $\gamma=1$, the purchase flow $dQ$ would vanish, because for $\gamma=1$, the characteristics of equation \eqref{eq:density} --- the trajectories of orders in the $(p,t)$ space --- would never intersect the market price line $A$. This means sellers would shift the prices of their orders by the exact same amount the market price shifts due to the impact of $u$, making execution impossible.
\end{remark}

\section{Formulation of the Optimal Execution Problem}

We consider the cost minimization problem:
\begin{equation}\label{eq:func_1}
J[u] := (1-\gamma) \int_0^T A(t)\, n(A(t), t)\, u(t)\, dt \;\to\; \min_{0 \le u\le U}
\end{equation}
for purchasing a volume $\bar Q > 0$ over a period $T>0$:
\begin{equation}\label{eq:constraints}
(1-\gamma) \int_0^T n(A(t), t)\, u(t)\, dt = \overline{Q}.   
\end{equation}
The dynamics of the order density are described by the transport equation \eqref{eq:density} with \eqref{eq:velocity}, and execution occurs at the market price level, governed by equation \eqref{eq:market_price} taking into account the long-term price dynamics \eqref{eq:f_price}.

\section{Necessary Optimality Conditions}

First, we reduce the system dynamics to a finite-dimensional case. To do this, note that by Corollary \ref{co:nAt}, the characteristic starting at $(0,p_0)$ has the form:
$$
p(t;\, p_0) = p_0 + A(t)- D(t), \quad D(t):=A_0+\int_0^t (1-\gamma)\, u(s)\, ds.
$$
An order with initial price $p_0$ is executed at the moment when $p(t;\,p_0) = A(t)$, that is, when $p_0 = D(t)$. Therefore, $n(A(t),t) =n_0(D(t))$, and the cost functional and the volume constraint can be written as:
\begin{equation}\label{eq:functional_n_0}
\mathcal{J}_0:=(1-\gamma)\int_0^T A(t)\, n_0(D(t))\, u(t)\, dt \to \min_{0\le u\le U},
\end{equation}
\begin{equation}\label{eq:limit_order_n_0}
\mathcal{J}_1 := (1-\gamma)\int_0^T n_0(D(t))\, u(t)\, dt - \overline{Q} = 0,
\end{equation}
using the additional phase variable $D$:
\begin{equation}\label{eq:D_equation}
\frac{dD}{dt}=(1-\gamma)\, u, \qquad D(0)=A_0.
\end{equation}

Hereafter, we additionally assume that the required volume is strictly less than the maximum possible:
\begin{equation}\label{eq:intQ}
0<\bar Q < \left(1-\gamma\right) U\int_0^T n_0(A_0+\left(1-\gamma\right)U\,t)\, dt,
\end{equation}
since the equality case would be trivial, as the admissible control would be unique ($u \equiv U$).

The Pontryagin Hamiltonian for the problem \eqref{eq:market_price}, \eqref{eq:f_price}, \eqref{eq:D_equation}, \eqref{eq:functional_n_0}, \eqref{eq:limit_order_n_0} takes the form:
\begin{equation}\label{eq:H}
  H=\xi\left(u-\rho(A-V)+b\,V\right)+\eta\left(\kappa\, u+b\,V\right)
+\zeta(1-\gamma)u-(\lambda\, A+\mu)(1-\gamma)n_0(D)\,u,
\end{equation}
where $\lambda$, $\mu$, $\xi$, $\eta$, and $\zeta$ are Lagrange multipliers.

According to the Pontryagin Maximum Principle for a problem with an isoperimetric constraint \cite[\S 4.2.2]{Tikhomirov_OC}, if $(\hat{A}, \hat{V},\hat{D},\hat{u})$ is an optimal process, then there exist multipliers:
$$
\hat{\lambda}\geq 0, \ \hat{\mu}, \ \hat{\xi}, \ \hat{\eta}, \ \hat{\zeta},
$$
not all simultaneously zero, consisting of constants $\hat{\lambda},$ $\hat{\mu}$ and adjoint variables $\hat{\xi}, \hat{\eta},$ $\hat{\zeta}$, such that:

a) the adjoint system and transversality conditions are satisfied:
\begin{align}
\dot{\xi} = -H'_{A} & = \lambda (1-\gamma) n_0(D)\, u + \rho\, \xi, & \xi(T) =0,\label{eq:xi_adj}\\[5pt]
\dot{\eta}= -H'_V & = -(b+\rho)\xi - b\,\eta, & \eta(T) = 0,\label{eq:eta}\\[3pt]
\dot{\zeta} = -H'_D & = (\lambda\, A+\mu)(1-\gamma) n'_0(D)\, u,  & \zeta(T)=0; \label{eq:zeta}
\end{align}

b) the maximality condition with respect to $u$ holds for almost all $t\in[0,T]$:
$$
H\big(t,\hat{A}(t),\hat{V}(t), \hat{D}(t),\hat\xi(t),\hat\eta(t),\hat\zeta(t),\hat u(t)\big)
=\max_{\tilde{u}\in[0,U]}H\big(t,\hat{A}(t),\hat{V}(t), \hat{D}(t),\hat\xi(t),\hat\eta(t),\hat\zeta(t),\tilde{u}\big);
$$

c) along with the complementary slackness condition:
$$
\mu \mathcal{J}_1 = 0.
$$

The linearity of \eqref{eq:H}, $H=\Phi\,u+\xi(bV-\rho(A-V))+\eta\, b\,V$, with respect to the control $u$, where
$$
\Phi:=\xi+\kappa\eta+(1-\gamma)\zeta-(1-\gamma)(\lambda A+\mu)n_0(D)
$$
is the switching function, determines the bang-bang structure of the optimal control $\hat u$ as follows:
$\hat u(t)=U$ when $\Phi(t)>0$, $\hat u(t)=0$ when $\Phi(t)<0$, and $\hat u(t)\in[0,U]$ when $\Phi(t)=0$.

\paragraph{Normality of the problem.}
Let $\lambda=0$. Then the adjoint system \eqref{eq:xi_adj}--\eqref{eq:zeta} implies $\xi\equiv\eta\equiv 0$ and
$
    \zeta(t) = \mu\big[n_0(D(t)) - n_0(D(T))\big].
$
If $\mu\neq 0$, then, since $n_0>0$ and $\zeta(T)=0$, we obtain
$$
\Phi(t) = -(1-\gamma)\,\mu\, n_0(D(T)) \neq 0,
$$
so $\Phi$ is a nonzero constant, and the maximum principle yields $\hat{u}\equiv 0$ or $\hat{u}\equiv U$, which contradicts \eqref{eq:intQ}. Hence $\mu=0$, but then all the multipliers vanish, which is prohibited. Therefore $\lambda>0$, and we normalize it by setting $\lambda=1$.

\subsection{Existence of Optimal Control}

We define the set of admissible controls, taking into account \eqref{eq:D_equation}, as:
$$
\mathcal{U} = \left\{u\in L^{\infty}(0,T)\ : (1-\gamma)\int_0^Tn_0(D(t))\,u(t)\,dt=\overline{Q},\ \ 0\le u\le U \text{ a.e.} \right\}.
$$

\begin{thrm}
Let $n_0\in C([A_0,\infty))$ be bounded, $n_0\ge0$, and the set of admissible controls be non-empty. Then the optimal execution problem has an optimal control $u^*\in \mathcal{U}$.
\end{thrm}
\begin{remark}
    This statement follows from general existence results \cite{Cesari1983}. For the sake of self-containment, we provide a proof using only standard facts from functional analysis.
\end{remark}

\begin{proof} 
For convenience, the optimal control problem is reduced to a Mayer problem:
$$
    J[u] := C(T) \to \min_{u\in\mathcal{U}}
$$
with the terminal constraint $Q(T) = \overline{Q}$, where $Q$ and $C$ satisfy the equations:
$$
    \dot{Q}(t) = (1-\gamma)n_0(D(t))\,u(t), \quad Q(0) = 0,
$$
$$
    \dot{C}(t) = (1-\gamma)A(t)\,n_0(D(t))\,u(t),\quad C(0)=0.
$$ 
Here, the set of admissible controls is $\mathcal{U} = \{ u\in L^{\infty}(0,T): 0\leq u\leq U \text{ a.e.}, Q(T)=\overline{Q}\}$.

Using the notation $x=(A,V,D,Q,C)$, the differential constraints are written as:
$$
    \dot{x}(t) = a(x(t)) + u(t)\, c(x(t)),\qquad x(0) = x_0 :=(A_0,V_0,A_0,0,0)
$$
with continuous vector functions:
$$
    a(x) = \big(-\rho(A-V)+b\,V,\ b\,V,\ 0,\ 0,\ 0\big),
$$
$$
    c(x) = \big(1, \kappa, 1-\gamma, (1-\gamma)n_0(D), (1-\gamma)A\,n_0(D)\big). 
$$

The linearity of $a$ with respect to $x$ and the boundedness of $n_0$ yield a linear growth estimate for any admissible control $u$ and the corresponding trajectory $x$:
$$||\dot{x}||\le M ||x|| + L \quad \text{for a.e.} \ t\in[0,T]$$
for some positive constants $L$ and $M$. Integrating this inequality and applying Gronwall's lemma to the function $t\mapsto ||x(t)||$, we obtain a uniform estimate $\|x(t)\|\le K$ for all $t\in[0,T]$, independent of the control: the family of admissible trajectories is uniformly bounded. Substituting this estimate into the linear growth inequality yields $\|\dot{x}(t)\|\le K_1$ for some new positive constant $K_1$ for almost all $t\in[0,T]$, and consequently, all admissible trajectories form an equicontinuous family.

Since all admissible trajectories on the non-empty set $\mathcal{U}$ satisfy $\| x(t)\|\leq K$ and the function $n_0(p)$ is bounded by some constant $N_0$, the functional $J[u]$ has a finite infimum. Hence, there exists a minimizing sequence $\{u_k \}_{k=1}^\infty\subset\mathcal{U}$ with $J[u_k]\to \inf_{u\in\mathcal{U}}{J}$. By the Arzela-Ascoli theorem \cite{KolmogorovFomin2009}, from the equicontinuous and uniformly bounded family of admissible trajectories $\{x_k\}$ corresponding to $\{u_k\}$, one can extract a subsequence converging uniformly on $[0,T]$ to some continuous $x^{*}$. Since all trajectories $x_k$ are Lipschitz continuous with a common constant $K_1$, their limit $x^*$ is also Lipschitz, and therefore absolutely continuous.

We show that the limit pair $(u^*,x^*)$ is admissible, i.e., in
$$
    x_k(t) = x_0 + \int_0^t a(x_k(s))ds + \int_0^t u_k(s) c(x_k(s)) ds,
$$
one can pass to the limit under the integral sign. For the first integral, the uniform convergence $x_k\to x^*$ and the continuity of $a$ immediately imply the uniform convergence $a(x_k)\to a(x^*)$. The second integral can be represented as:
\begin{equation}\label{eq:proof_ex_1}
    \int_0^t u_k(s) c(x^*(s))ds + \int_0^t u_k(s) \big(c(x_k(s))-c(x^*(s))\big) ds.
\end{equation}
We show that the first term in \eqref{eq:proof_ex_1} converges to $\int_0^t u^*(s)c(x^*(s))ds$ as $k\to\infty$. The family of primitives of the controls $\psi_k(t):=\int_0^t u_k(s) ds$ is equicontinuous and uniformly bounded; by the Arzela-Ascoli theorem \cite{KolmogorovFomin2009}, we extract a subsequence $\psi_k\to \psi$ (re-indexing $k$) converging uniformly on $[0,T]$, while the uniform convergence $x_k\to x^*$ is preserved. Each $\psi_k$ is non-decreasing and Lipschitz with constant $U$; the limit function inherits both properties. Consequently, by Lebesgue's theorem, the derivative of the limit of the primitives $u^*=\psi'$ exists a.e. and satisfies $0\le u^*(t)\le U$ a.e., meaning $u^*\in L^{\infty}(0,T)$.

Since $c(x^*(s))$ is continuous on $[0,T]$, and $\psi_k$ is non-decreasing and converges uniformly to $\psi$ on $[0,T]$, the passage to the limit in the Riemann-Stieltjes integral is valid (Helly's selection theorem \cite{KolmogorovFomin2009}):
$$
\int_0^t u_k(s) c(x^*(s)) ds = \int_0^t c(x^*(s)) d\psi_k(s) \rightarrow \int_0^tc(x^*(s)) d\psi(s).
$$
Since $\psi'=u^*$ and $\psi$ is absolutely continuous, it follows that:
$$
\int_0^t c(x^*(s)) d\psi(s) =\int_0^t u^*(s) c(x^*(s)) ds.
$$

The second term in \eqref{eq:proof_ex_1} tends to zero, since:
$$
\left| \int_0^t u_k(s) \big(c(x_k(s))-c(x^*(s))\big) ds\right|\leq U \cdot T \cdot \|c(x_k)-c(x^*) \|_{C([0,T])}\to 0,
$$
where $\|c(x_k)-c(x^*) \|_{C([0,T])} \to 0$ due to the uniform convergence $x_k\to x^*$ and the uniform continuity of $c$ on the ball $\|x \|\le K$.

Thus, the limit function $x^*$ satisfies the equation:
$$
    x^*(t) = x_0 + \int_0^t \big(a(x^*(s))+u^*(s) c(x^*(s)) \big) ds.
$$
Since $x_k\to x^*$ uniformly on $[0,T]$, we have $x_k(T)\to x^*(T)$, and $Q_k(T)=\overline{Q}$ holds for all $k$. Passing to the limit yields $Q^*(T)=\overline{Q}$.

From the uniform convergence $C_k(T)\to C^*(T)$ and the definition of the minimizing sequence $C_k(T)=J[u_k]\to \inf J$, we obtain $J[u^*] = \inf J$. Hence, $u^*$ attains the minimum of the functional.
\end{proof}

\section{Conclusion}

A fundamental feature of the proposed model is the endogenous dynamics of the order book shape: the book density evolves over time, reflecting the sellers' reaction to purchases. Models with a similar formulation were considered in \cite{ObizhaevaWang2013} and \cite{AlfonsiFruthSchied2010}. In the first, the order density is assumed to be block-shaped and constant; in the second, an arbitrary book shape is allowed, but the density function is fixed in time. In fact, in these models, only the position of the boundary where order execution occurs changes. In contrast to these approaches, the proposed model describes the evolution of the order distribution across prices during execution, making the shape of the order book an internal characteristic of the market dynamics rather than a predefined external parameter.

\end{document}